\documentstyle[11pt]{article}
\def\pl{ \: + }

\newfont{\bbbold}{msbm10 scaled \magstep1}

\def\cA{\cal A}

\def\cE{\cal E}
\def\cF{{\cal F}}

\def\cL{\cal L}

\newfont{\goth}{eufm10 scaled \magstep1}

\def\a{\alpha}
\def\b{\beta}
\def\c{\gamma}\def\C{\Gamma}
\def\d{\delta}
\def\e{\epsilon}

\def\h{\eta}
\def\k{\kappa}
\def\l{\lambda}\def\L{\Lambda}
\def\m{\mu}

\def\P{\Pi}

\def\s{\sigma}

\def\th{\theta}\def\Th{\Theta}

\def\be{\begin{equation}}\def\ee{\end{equation}}
\def\bea{\begin{eqnarray}}\def\eea{\end{eqnarray}}
\def\ba{\begin{array}}\def\ea{\end{array}}

\def\del{\partial}
\def\ua{\underline{\alpha}}
\def\ub{\underline{\phantom{\alpha}}\!\!\!\beta}
\def\uc{\underline{\phantom{\alpha}}\!\!\!\gamma}

\def\una{\underline a}\def\unA{\underline A}
\def\unb{\underline b}\def\unB{\underline B}
\def\unc{\underline c}\def\unC{\underline C}

\def\unm{\underline m}\def\unM{\underline M}

\def\unG{\underline{G}}
\def\unT{\underline{T}}

\def\xz{\times}

\def\nab{\nabla}

\def\del{\partial}

\let\la=\label

\let\bm=\bibitem

\def\umu{{\underline \mu}}
\def\nn{\nonumber}
\def\bd{\begin{document}}
\def\ed{\end{document}}
\def\bea{\begin{eqnarray}}
\def\eea{\end{eqnarray}}
\def\ft#1#2{{\textstyle{{\scriptstyle #1}\over {\scriptstyle #2}}}}
\def\fft#1#2{{#1 \over #2}}
\newcommand{\eq}[1]{(\ref{#1})}
\def\eqs#1#2{(\ref{#1}-\ref{#2})}
\def\det{{\rm det\,}}
\def\tr{{\rm tr}}
\newcommand{\ho}[1]{$\, ^{#1}$}
\newcommand{\hoch}[1]{$\, ^{#1}$}
\newcommand{\tamphys}{\it\small Center for Theoretical Physics,
Texas A\&M University, College Station, TX 77843, USA}
\newcommand{\newton}{\it\small Isaac Newton Institute for Mathematical
Sciences, Cambridge, UK}
\newcommand{\kings}{\it\small Department of Mathematics, King's College,
London, UK}

\newcommand{\auth}{\large P.S. Howe\hoch{1}, O.
Raetzel\hoch{1}\hoch{\dagger} I. Rudychev \hoch{2} and E. Sezgin
\hoch{2} }

\begin{document}

\hfill{KCL-MTH-98-39}

\hfill{CTP-TAMU-34/98}

\hfill{hep-th/9810081}

\hfill{\today}

\vspace{20pt}

\begin{center}

{\Large{\bf L-branes}}

\vspace{30pt}

\auth

\vspace{15pt}

\begin{itemize}
\item [$^1$] \kings
\item [$^2$] \tamphys
\end{itemize}

\vspace{60pt}

{\bf Abstract}

\end{center}

The superembedding approach to $p$-branes is used to study a class of
$p$-branes which have linear multiplets on the worldvolume. We refer to
these branes as L-branes. Although linear multiplets are related to
scalar multiplets (with 4 or 8 supersymmetries) by dualising one of the
scalars of the latter to a $p$-form field strength, in many geometrical
situations it is the linear multiplet version which arises naturally.
Furthermore, in the case of 8 supersymmetries, the linear multiplet is
off-shell in contrast to the scalar multiplet. The dynamics of an
L-brane are obtained by using a systematic procedure for constructing
the Green-Schwarz action from the superembedding formalism. This action
has a Dirac-Born-Infeld type structure for the $p$-form. In addition, a
set of equations of motion is postulated directly in superspace, and is
shown to agree with the Green-Schwarz equations of motion.

{\vfill\leftline{}\vfill
\vskip  10pt
\footnoterule
{\footnotesize
\hoch{\dagger} E-mail: oraetzel@mth.kcl.ac.uk
}

\pagebreak
\setcounter{page}{1}

\section{Introduction}

In recent years there has been renewed interest in superstring theory as
a candidate theory that unifies all the fundamental forces in nature.
The interest was sparked by the realisation that the five different
consistent ten-dimensional superstring theories are in fact related to
each other by duality transformations. Furthermore, it is believed that
they are related to a new theory in eleven dimension which has been
called M-theory. A crucial r\^{o}le in this development has been played
by the soliton solutions of superstrings and eleven-dimensional
supergravity which correspond geometrically to multi-dimensional objects
present in the spectra of these theories. These are generically referred
to as $p$-branes.

One way of studying the dynamics of $p$-branes is to use the theory of
superembeddings \cite{se1, se2, hs1}. In this framework the worldvolume
swept out by a $p$-brane is considered to be a subsupermanifold of a
target superspace. A natural restriction on the embedding gives rise to
equations which determine the structure of the worldvolume
supermultiplet of the $p$-brane under consideration and which may also
determine the dynamics of the brane itself. There are several types of
worldvolume supermultiplets that can arise: scalar multiplets, vector
multiplets, tensor multiplets which have 2-form gauge fields with
self-dual field strengths, and multiplets with rank 2 or higher
antisymmetric tensor gauge fields whose field strengths are not
self-dual \cite{hs1}. Although this last class of multiplets can be
obtained from scalar or vector multiplets by dualisation (at least at
the linearised level) it is often the case that the version of the
multiplet with a higher rank gauge field is more natural in a given
geometrical context. For example, D-branes in type II string theory have
worldvolume vector multiplets but, in the context of D$p$-branes ending
on D$(p+2)$-branes, the dual multiplet of the latter brane occurs
naturally \cite{chs1}.

In this paper we discuss a class of $p$-branes whose members are
referred to as L-branes. By definition, these branes are those which
have worldvolume supermultiplets with higher rank non-self-dual tensor
gauge fields which are usually referred to as linear multiplets, whence
the appellation. According to \cite{hs1} there are two sequences of
L-branes: the first has as its members a 5-brane in $D=9$, a 4-brane in
$D=8$ and a 3-brane in $D=7$ which all have eight worldvolume
supersymmetries, while the second has only one member, the 3-brane in
$D=5$ which has four worldvolume supersymmetries. These sequences and
their worldvolume bosonic field contents are tabulated below, where $A_p$
denotes a $p$-form potential and $(S,T)$ are auxiliary fields.

\begin{center}
\begin{tabular}
{|c|c|c|l|}
\hline
8 world susy & \ \ L5-brane  & \ \  D=9 &\ \ $3\phi, \ A_4$  \\
         & \ \ L4-brane  & \ \  D=8 &\ \ $3\phi, \ A_3, S $  \\
         & \ \ L3-brane  & \ \  D=7 &\ \ $3\phi, \ A_2, S,T$  \\
\hline
4 world susy  & \ \ L3-brane  & \ \  D=5 &\ \ $\phi,  \ A_2$  \\
\hline
\end{tabular}
\end{center}

The 3 and 4-branes of the first sequence can be obtained by double
dimensional reduction from the first member of the sequence, namely the
5-brane in $D=9$, and the latter can be intepreted as arising as a
vertical reduction of the geometrical sector of the heterotic/type I
\linebreak
5-brane, followed by the dualisation of the scalar field in the
compactified direction. By the geometrical sector we mean the sector
containing the worldvolume fields corresponding to the breaking of
supertranslations. The relevant target space field theory in this
context is the dimensional reduction of the dual formulation of $N=1,
D=10$ supergravity followed by a truncation of a vector multiplet. The
L5-brane is expected to arise as a soliton in this theory. Another
possible interpretation of this brane is as the boundary of a D6-brane
ending on a D8-brane. In this case the target space geometry should be
the one induced from the embedding of the D8-brane in type IIA
superspace. However, in this paper we shall take the target space to be
flat $N=1, D=9$ superspace for simplicity. The generalisation to a
non-trivial supergravity background is straightforward while the
generalisation to an induced D8-brane background geometry is more
complicated and would require further investigation.

The L3-brane in $D=5$ also has two possible interpretations. On the one
hand it may arise as the geometrical sector of a soliton in
heterotic/type I theory compactified on $K3$ to $D=6$, vertically reduce
to $D=5$ and then dualised in the compactified direction. Alternatively,
it could be related to the triple intersection of D-branes \cite{bdr}
over a 3-brane with one overall transverse direction. This brane will be
studied in \cite{cod1} as an example of a brane of codimension one.

A feature of L-branes is that their worldvolume multiplets are off-shell
multiplets in contrast to many of the branes that have been studied
previously such as M-branes and D-branes. In particular, this is true
for the worldvolume multiplets of the first sequence of L-branes, even
though their dual versions involve hypermultiplets which are unavoidably
on-shell. The standard embedding constraint does not lead to the
dynamics of L-branes and imposing the Bianchi identity for the
worldvolume tensor gauge field does not change the situation. As a
consequence the equations of motion of such branes have to be determined
by other means, either by directly imposing an additional constraint in
superspace or by using the recently proposed brane action principle
which has the advantage of generating the modified Born-Infeld term for
the tensor gauge fields in a systematic way \cite{hos}. We note that the
heterotic 5-brane in $N=1, D=10$, which is related to the L5-brane in $D=9$ as
we described above, would normally be described by a worldvolume hypermulitplet
which is on-shell. However, as noted in \cite{hs1}, one can go off-shell using
harmonic superspace methods. This has been discussed in detail in a recent
paper \cite{bik}.

In the foregoing discussion we have assumed throughout that the target space
supersymmetry is minimal. It is possible to relax this. For example, one can
obtain an L2-brane  in an $N=2, D=4$ target space by double dimensional
reduction of the L3-brane in $D=5$. This brane and the non-linear dynamics of
the associated linear multiplet has been studied from the point of view of
partial breaking of supersymmetry in reference \cite{bg}.

The organisation of the paper is as follows: in the next section we give
a brief introduction to the theory of superembeddings; in section 3 we study
the L5-brane in $D=9$ at the linearised level; in section 4 the torsion
and Bianchi identities are solved in the non-linear theory; in section 5
\linebreak
we construct the action and in section 6 we derive the Green-Schwarz
equations of motion and determine how these can be expressed in
superspace. In section 7 the L3-brane in $D=7$ and L4-brane in $D=8$ are
studied. Some concluding remarks are made in section 8. Our conventions
for spinors, gamma matrices and superforms are given in an appendix.

%%%%%%%%%%%%%%%%%%%%%%%%%%%%%%%%%%%%%%%%%%%%%%%%%%%%%%%%%%%%%%%%%%%%%%%%

\section{Superembeddings}

%%%%%%%%%%%%%%%%%%%%%%%%%%%%%%%%%%%%%%%%%%%%%%%%%%%%%%%%%%%%%%%%%%%%%%%%

In the superembedding approach to $p$-branes both the target
space and the worldvolume swept out by the brane are superspaces. This
is different to the Green-Schwarz formalism where only the target space
is taken to be a superspace while the worldvolume is purely bosonic. The
local $\k$-symmetry of the GS formalism can be understood as arising
from the local supersymmetry of the worldvolume in the superembedding
approach upon gauge-fixing.

The geometric principles underlying the superembedding approach were
given in \cite{hsw}. The embedding $f:M\rightarrow \unM$, which maps the
worldvolume $M$ into the target superspace $\unM$, is chosen to break
half of the target space supersymmetries so that the fermionic dimension
of $M$ must be chosen to be half the fermionic dimension of $\unM$. More
general embeddings which break more of the supersymmetries are possible
but will not be considered here. We adopt the general convention that
worldvolume quantities are distinguished from target space quantities by
underlining the latter. Coordinates on the worldvolume (target space)
are denoted by $z^M=(x^m,\th^{\m})$ and $z^{\unM} = ({x}^{\unm} ,
{\th}^{\umu})$ respectively; the tangent bundles are denoted by $T$
($\unT$) and local preferred bases are denoted $E_A=(E_a,E_{\a})$ and
$E_{\unA} = (E_{\una} , E_{\ua}$). The associated cotangent bundles
$T^*$ ($\unT^*$) are similarly spanned by the dual basis one-forms
$E^A=(E^a, E^{\a})$ and $E^{\unA} = (E^{\una}, E^{\ua}$). The target
space supervielbein $E_{\unM}{}^{\unA}$ and its inverse
$E_{\unA}{}^{\unM}$ are used to change from a preferred basis to a
coordinate basis. The worldvolume supervielbein and its inverse are
similarly denoted, but without underlining of the indices. In the
foregoing latin indices are bosonic and greek indices are fermionic.

The embedding matrix $E_A{}^{\unA}$ specifies the relationship between
the bases on $T$ and $\unT$

\be
E_A = E_A{}^{\unA}E_{\unA}\ .
\label{1}
\ee

Expressed in local coordinates the embedding matrix is given by

\be
E_A{}^{\unA} = E_A{}^{M}\del_{M}z^{\unM}E_{\unM}{}^{\unA}\ .
\label{2}
\ee

The basic embedding condition is that the purely fermionic part of $T$
is determined only by the pullback of the purely fermionic part of
$\unT$ and does not involve the pullback of the bosonic part of $\unT$.
This means that the embedding matrix should satisfy the constraint

\be
E_{\a}{}^{\una} = 0\ .
\label{3}
\ee

This basic embedding constraint determines the supermultiplet structure
of the brane and in many cases will also be enough to put the brane
on-shell. In some cases, however, an additional constraint involving
forms on the worldvolume and the target space is necessary. The
L-branes discussed in this paper are particularly interesting in this
regard because the embedding constraint \eq{3} is not sufficient to
determine the dynamics of these branes so that the equations of motion
must be derived from additional constraints or actions.

In order to make further progress it is convenient to introduce the
normal tangent bundle $T'$ which has a basis denoted by $E_{A'}=(E_{a'},
E_{\a'})$. This basis is related to the basis of $\unT$ by the normal
matrix $E_{A'}{}^{\unA}$. Note that normal indices are distinguished from
tangent indices by primes. There is considerable freedom in the choice
of the components of $E_{A}{}^{\unA}$ and $E_{A'}{}^{\unA}$. A simple and
convenient choice is

\be
\ba{rclrcl}
E_{a}{}^{\una} & = & u_{a}{}^{\una} &\qquad\qquad
E_{\a}{}^{\una}& = & 0 \\
E_{a}{}^{\ua} &= &\L_{a}{}^{\b'}u_{\b'}{}^{\ua}&
E_{\a}{}^{\ua} & = & u_{\a}{}^{\ua} + h_{\a}{}^{\b'}u_{\b'}{}^{\ua}\\
&&&&&\\
E_{a'}{}^{\una}& = & u_{a'}{}^{\una}&\qquad\qquad
E_{\a'}{}^{\una}&=&0 \\
E_{a'}{}^{\ua}&=&0&
E_{\a'}{}^{\ua} & = & u_{\a'}{}^{\ua}\ .
\ea
\label{4}
\ee

where $\L_{a}{}^{\b'}$ is an arbitrary vector-spinor. The matrices
$u_{a}{}^{\una}$ and $u_{a'}{}^{\una}$ together make up an element of
the Lorentz group of the target space and, similarly, the matrices
$u_{\a}{}^{\ua}$ and $u_{\a'}{}^{\ua}$ make up the corresponding element
of the spin group.
With this choice the components of the inverse matrices
$((E^{-1})_{\unA}{}^{A}, (E^{-1})_{\unA}{}^{A'})$ are given by

\be
\ba{rclrcl}
(E^{-1})_{\una}{}^{a} & = & (u^{-1})_{\una}{}^{a} &\qquad\qquad
(E^{-1})_{\una}{}^{\a}& = & 0 \\
(E^{-1})_{\ua}{}^{a} &= &0   &
(E^{-1})_{\ua}{}^{\a} & = & (u^{-1})_{\ua}{}^{\a}\\
&&&&&\\
(E^{-1})_{\una}{}^{a'} & = & (u^{-1})_{\una}{}^{a'}&\qquad\qquad
(E^{-1})_{\una}{}^{\a'} & = & -(u^{-1})_{\una}{}^{b}\L_{b}{}^{\a'}\\
(E^{-1})_{\ua}{}^{a'}&=&0&
(E^{-1})_{\ua}{}^{\a'} & = & (u^{-1})_{\ua}{}^{\a'} -
(u^{-1})_{\ua}{}^{\b}h_{\b}{}^{\a'}\ .
\ea
\label{6}
\ee

The consequences of the basic embedding condition can be conveniently
analysed by means of the torsion identity which is simply the equation
defining the target space torsion tensor pulled back onto the
worldvolume by means of the embedding matrix. Explicitly, one has

\be
\nabla_A E_B{}^{\unC}-(-1)^{AB}\nabla_B E_A{}^{\unC}+T_{AB}{}^C E_C{}^{\unC}
= (-1)^{A(B+\unB)}E_B{}^{\unB}E_A{}^{\unA}T_{\unA \unB}{}^{\unC}.
\label{25}
\ee

where $\nabla$ denotes a covariant derivative which acts independently
on the target space and worldvolume indices. There are different
possibilites for this derivative and we shall specify our choice later.
With the embedding condition \eq{3} the dimension zero component of the
torsion identity does not involve any connection terms and reduces to
the simple form

\be
T_{\a \b}{}^{c}E_{c}{}^{\unc} = E_{\a}{}^{\ua}E_{\b}{}^{\ub}
T_{\underline{\a \b}}{}^{\unc}
\label{12}
\ee

In a sense one can consider the dimension zero component of the torsion
tensor (Frobenius tensor) as the basic tensor in superspace geometry and
the above equation specifies how the target space and worldvolume
Frobenius tensors are related.

The flat target superspace geometry for L$p$-branes $(p=3,4,5)$ also
includes the following differential forms

\bea
G_2 &=& dC_1\ , \nn\\
G_{p+1} & = & dC_p\ ,\nn\\
G_{p+2} &=& dC_{p+1}-C_1 G_{p+1} \ ,  \qquad p=3,4,5,
\eea

which obey the Bianchi identities

\bea
dG_2 &=& 0\ , \nn\\
dG_{p+1} &=& 0\ ,\nn\\
dG_{p+2} &=& G_2 G_{p+1}\ . \la{ggg}
\eea

In the flat target space under consideration here the non-vanishing
components of the forms $G_q$ are

\be
G_{\ua \ub \una_{1} \ldots \una_{(q-2)}} = -i(\C_{\una_{1}
\ldots \una_{(q-2)}})_{\ua \ub}\ ,
\label{10}
\ee

except for $G_{\ua \ub \una\unb\unc} $ which arises for the L$4$-brane in
$D=8$, in which case a factor of $\C_9$ is needed so that
$(\C_{\una\unb\unc}\C_9)_{\ua\ub}$ has the right symmetry. In $D=8,9$
the spinor indices label $16$ component pseudo-Majorana spinors while in
$D=7$ they represent a pair of indices, one of which is an $Sp(1)$
doublet index, which together label a $16$ component symplectic-Majorana
spinor.

In proving the Bianchi identities $dG_{p+1}=0$ for $Lp$-branes in
$(p+4)$ dimensions, the following $\C$-matrix identities are needed:

\be
(\C_{\una_1})_{(\underline{\a\b}}\,(\C^{\una_1 \cdots
\una_{(p-1)}}){}_{\underline{\c\d})}=0\ . \la{ci}
\ee

These identities are well known in the context of the usual
$p=2,3,4,5$-branes in $D=7,8,9,10$, respectively \cite{bst,act}. To
prove the Bianchi identity $dG_{p+2}=G_2G_{p+1}$ for an L$p$-brane in
$(p+4)$ dimensions, on the other hand, one needs to use a $\C$-matrix
identity resulting from the dimensional reduction of \eq{ci} from one
dimension higher. For example, to prove $dG_7=G_2G_6$ for the L$5$-brane
in $D=9$, one needs the following identity

\be
(\C_{\una_1})_{(\underline{\a\b}}\,(\C^{\una_1 \cdots
\una_5}){}_{\underline{\c\d})}
+ C_{(\underline{\a\b}}\,(\C^{\una_2\cdots \una_5})_{\underline{\c\d})}=0\
,\la{ci2}
\ee

which follows from the identity \eq{ci}, which holds in $D=10$,
by a dimensional reduction to $D=9$.

In addition to the geometrical quantities for each L$p$-brane there is a
$(p-1)$-form worldvolume gauge field ${\cA}_{p-1}$ with modified field
strength $p$-form ${\cF}_{p}$ defined by

\be
{\cF}_p = d{\cA}_{p-1}- \unC_p\ ,\qquad p=3,4,5\ ,
\ee

where $\unC_{p}$ is the pull-back of a target space $p$-form
$C_{p}$. This field stength obeys the Bianchi identity

\be
d{\cF}_p=-\unG_{p+1}\ ,
\ee

where $\unG_{p+1}$ is the pull-back of a target space $(p+1)$-form
$G_{p+1}$. In the first sequence of L$p$-branes this identity is a
consequence of the basic embedding condition \eq{3}, while for the
L3-brane in $D=5$ this is not the case. For this brane the ${\cF}$
Bianchi identity is required in order to completely specify the
worldvolume supermultiplet as a linear multiplet.

%%%%%%%%%%%%%%%%%%%%%%%%%%%%%%%%%%%%%%%%%%%%%%%%%%%%%%%%%%%%%%%%%%%%%%

\section{The Linearized Theory}

%%%%%%%%%%%%%%%%%%%%%%%%%%%%%%%%%%%%%%%%%%%%%%%%%%%%%%%%%%%%%%%%%%%%%

Let us consider the linearisation of a flat brane in a flat target
space. The target space basis forms are

\bea
E^{\una} &= & dx^{\una}
- {i\over 2} d\th^{\ua}(\C^{\una})_{\ua\ub}\th^{\ub}\ , \nn\\
E^{\ua} &=& d\th^{\ua}\ .
\la{Ealfa}
\eea

In the physical gauge we have

\bea
x^{\ua} &= &(x^a,x^{a'}(x,\th){})\ ,\nn\\
\th^{\ua} &= &(\th^{\a},\Th^{\a'}(x,\th){})\ .
\la{theta}
\eea

where it is supposed that the fluctuations of the brane which are
described by the transverse (primed) coordinates as functions of the
(unprimed) brane coordinates are small. In addition we write the
worldvolume basis vector fields in the form $E_A{}^M\del_M = D_A-H_A{}^B
D_B$ where $D_A = (\del_a,D_{\a})$ is the flat covariant derivative on
the worldvolume. In the linearised limit, the elements of the embedding
matrix take the form

\be
E_a{}^{\unb} = (\d_a{}^b{},{}\del_a X^{b'}),
\la{E1}
\ee

\be
E_{\a}{}^{\ub} = (\d_{\a}{}^{\b}{},{}D_{\a}\Th^{\b'}),
\la{E2}
\ee

\be
E_a{}^{\ub} = ({}0{},{}\del_a \Th^{\b'}).
\la{E3}
\ee

Consequently, \eq{3} tells us that the deformation $H_A{}^B$ of the
supervielbein vanishes and that

\be
D_{\a} X^{a'} = i(\C^{a'})_{\a\b'}\Th^{\b'},
\la{lin basic}
\ee

where

\be
X^{a'} = x^{a'} \pl {i\over 2} \th^{\a}
(\C^{a'})_{\a\b'}\Th^{\b'}.
\la{X}
\ee

For the  L-branes of the first sequence \eq{lin basic} takes the form

\be
D_{\a i} X^{a'} = i(\c^{a'})_{ij} \Th^j_{\a}\ , \qquad a'=1,2,3;\ i=1,2\ .
\la{basic L}
\ee

Defining

\be
X_{ij} = (\c^{a'})_{ij}X_{a'}\ ,
\ee

substituting into \eq{basic L} and multiplying by $\c_{a'}$
we get

\be
D_{\a i} X_{jk} =-i(\e_{ij}\Th_{\a k} \pl
\e_{ik}\Th_{\a j}).
\la{mult X}
\ee

This equation describes the linear multiplet with eight supersymmetries
in $d=6,5,4$ corresponding to the $L5$, $L4$ and $L3$-branes of the first
sequence. The field content of this multiplet consists of 3 scalars, an
8-component spinor and a divergence-free vector in all cases together
with an additional auxiliary scalar in $d=5$ and two auxiliary scalars
in $d=4$. The dual of the divergence-free vector can be solved for in
terms of a $4,3$ or 2-form potential in $d=6,5,4$ dimensions
respectively. For example, in $d=6$, spinorial differentiation of
\eq{mult X} gives

\be
D_{\a i}\Th_{\b j} = \e_{ij}(\c^a)_{\a\b}h_a
+\ft12  (\c^a)_{\a\b}\del_a X_{ij},
\la{DG}
\ee

where $h_a$ is the conserved vector in the multiplet, $\del^a h_a=0$.
This field, together with the 3 scalars $X_{ij}$ and the 8 spinors
$\Th_{\a i}$ (evaluated at $\th=0$) are the components of the
(off-shell) linear multiplet. At the linearised level the field
equations are obtained by imposing the free Dirac equation on the spinor
field $(\c^a )^{\a \b} \del_a \Th_{\b}{}^j = 0$. One then finds the
Klein-Gordon equation $\del_a \del^aX_{ij}=0$ for the scalars and the
field equation for the antisymmetric tensor gauge field $\del_{[a}
h_{b]}=0$.

%%%%%%%%%%%%%%%%%%%%%%%%%%%%%%%%%%%%%%%%%%%%%%%%%%%%%%%%%%%%%%%%%%%%%%

\section{The L5-brane in D=9}

%%%%%%%%%%%%%%%%%%%%%%%%%%%%%%%%%%%%%%%%%%%%%%%%%%%%%%%%%%%%%%%%%%%%%

We now turn to a detailed discussion of the L5-brane in $D=9$ in a flat
target superspace. We begin with a brief discussion of the target space
geometry. This can be derived most simply by dimensional reduction from
the flat $N=1, D=10$ supergeometry. In this way or by a direct
construction, one can establish, in addition to the usual supertorsion,
the existence of the $2,6,7$ forms $G_2,G_6,G_7$ defined in \eq{ggg}.

In $N=1,D=9$ flat superspace, the only non-vanishing component of the
torsion tensor is

\be
T_{\ua \ub}{}^{\unc}=-i(\C^{\unc})_{\ua\ub}\ .
\la{tt}
\ee

With this background we can now start to study the details of the
L5-brane using the torsion identity \eq{25}. From this starting point
there are now two equivalent ways to proceed. The first one is to fix
some of the components of the torsion tensor on the worldvolume
$T_{AB}{}^C$ in a convenient form. The second one is to fix the
connection used in \eq{25} by specifying some of the components of
the tensor $X_{A, B}{}^{C}$ defined by

\be
X_{A,B}{}^C \equiv (\nab_A u_B{}^{\unC})(u^{-1})_{\unC}{}^C\ .
\label{28}
\ee

Using this method all the components of the worldvolume torsion can be
found in terms of the vector-spinor $\Lambda_b{}^{\beta'}$ introduced in
\eq{4}. Although the two methods are equivalent, we have found that in
practice the second method is more efficient and will be used here. The
components of $X_{A, B}{}^{C}$ can be chosen to be

\begin{eqnarray}
X_{A,b}{}^c & = & X_{A,b'}{}^{c'} = 0 \nonumber \\
X_{A,\beta}{}^\gamma & = & X_{A,\beta'}{}^{\gamma'} = 0.
\label{29}
\end{eqnarray}

Note that, since $X_{AB}{}^C$ takes its values in the Lie algebra of the
target space Lorentz group, the components with mixed primed and
unprimed spinor indices are determined by the components with mixed
vector indices. Thus we have

\bea
X_{A,\b}{}^{\c'} &=& \ft14 (\C^{bc'})_{\b}{}^{\c'} X_{A,bc'}\ ,\nn\\
X_{A,\b'}{}^{\c} &=& \ft14 (\C^{bc'})_{\b'}{}^{\c} X_{A,bc'}\ .
\eea

We shall now analyse the torsion identity \eq{25} order by order in
dimension starting at dimension zero.

%%%%%%%%%%%%%%%%%%%   Dim 0 Torsion Story  %%%%%%%%%%%%%%%%%%%%%

{\bf Dimension 0}

We recall that the dimension 0 component of the torsion identity is

\be
T_{\a \b}{}^{c}E_{c}{}^{\unc} = E_{\a}{}^{\ua}E_{\b}{}^{\ub}
T_{\underline{\a \b}}{}^{\unc}
\label{12'}
\ee

Projecting \eq{12'} with $(E^{-1})_{\unc}{}^{c'}$ and using the expressions for
the embedding matrix given in \eq{4} we find

\be
h_{\a i \b}{}^{k}(\c^{c'})_{kj} + h_{\b j \a}{}^{k}(\c^{c'})_{ik} = 0\ ,
\label{15}
\ee

which can be solved for $h_{\a}{}^{\b'}$ to give

\be
h_{\a}{}^{\b'} = h_{\a i \b}{}^{j} =
\d_{i}{}^{j}h_{a}(\c^{a})_{\a \b}\ ,
\label{16}
\ee

where $h^2 \equiv h^a h_a$. Projection with $(E^{-1})_{\unc}{}^{c}$ on
the other hand yields

\be
T_{\a \b}{}^{a} = -i \e_{ij}(\c^{b})_{\a \b}m_{b}{}^a\ ,
\label{17}
\ee

where $m_{a}{}^{b}$ is given by

\be
m_{a}{}^{b} = (1-h^2)\d_{a}{}^{b} + 2h_{a}h^{b}\ .
\label{18}
\ee

At the linearised level the field $h_a$ coincides with the
divergence-free vector field in the linear multiplet discussed in the
previous section. As we shall see later the divergence-free condition
receives non-linear corrections in the full theory.

%%%%%%%%%%%%%%%  Dim  1/2 Torsion Story  %%%%%%%%%%%%%%%%%%%%%%%%%%%%

{\bf Dimension 1/2}

At dimension $1 \over 2$ equation \eq{25} gives rise to two equations

\be
\nab_\a E_b{}^{\unc}+T_{\a b}{}^c E_c{}^{\unc} = -E_b{}^{\ub}
E_\a{}^{\ua}T_{\underline{\a \b}}{}^{\unc}
\label{33}
\ee

and

\be
\nab_\a E_{\b}{}^{\uc} + \nab_\b E_\a{}^{\uc} + T_{\a \b}{}^{\c}
E_\c{}^{\uc} = -T_{\a\b}{}^c E_c{}^{\uc}.
\label{34}
\ee

Projection of \eq{33} with $(E^{-1})_{\unc}{}^c$ gives

\be
T_{\a b}{}^c = i h_a (\c^a\c^c)_{\a}{}^{\b} \L_{b \b i}\ ,
\label{36}
\ee

while projection  with $(E^{-1})_{\unc}{}^{c'}$  gives

\be
X_{\a i,b}{}^{c'} = -i(\c^{c'})_{ij} \L_{b \a}{}^j\ .
\label{37}
\ee

The dimension  ${1 \over 2}$ component of $X_{A,B}{}^C$,
$X_{\a,\b}{}^{\c'}$,  is then determined due to the fact that

\be
X_{\a,\b}{}^{\c'}
= -{1 \over 2}(\c^{b})_{\b \c}(\c^{c'})_{j}{}^{k}X_{\a i, bc'}\ .
\label{38}
\ee

Similarly one finds that

\be
X_{\a,\b'}{}^{\c}= {1 \over 2}(\c^{b})^{\b \c}(\c^{c'})_{j}{}^{k}X_{\a i, bc'}\
{}.
\ee

Projecting \eq{34} with $(E^{-1})_{\uc}{}^{\c}$ we find

\be
T_{\a \b}{}^{\c} = -h_{\b}{}^{\d'} X_{\a,\d'}{}^{\c}-h_{\a}{}^{\d'}
X_{\b,\d'}{}^{\c}\ ,
\label{39}
\ee

while projecting onto the transverse space with $(E^{-1})_{\uc}{}^{\c'}$ we
find

\be
\nab_{\a i} h_{a} = {i \over 2}(\tilde{\L}_{\a i a}
- (\c_{a}\c^{b})_{\a}{}^{\b} \tilde{\L}_{\b i b})\ ,
\label{40}
\ee

where $\tilde{\L}_{\a i b}$ is defined by

\be
\tilde{\L}_{\a i a} \equiv m_{a}{}^{b} \L_{\a i b}\ .
\label{41}
\ee

All dimension ${1 \over 2}$ quantities on the worldvolume can therefore
be expressed in terms of the vector-spinor $\L_{b}{}^{\b'}$ and the
worldvolume vector $h_a$.

%%%%%%%%%%%%% Dim 1 Torsion Story %%%%%%%%%%%%%%%%%

{\bf Dimension 1}

At dimension $1$ equation \eq{25}  again gives two equations but now
involving spacetime derivatives of the embedding matrix
$E_{A}{}^{\unA}$, namely

\be
\nab_a E_b{}^{\unc} - \nab_b E_a{}^{\unc} + T_{ab}{}^c E_c{}^{\unc}
= E_b{}^{\ub} E_a{}^{\ua} T_{\underline{\a \b}}{}^{\unc}
\label{42}
\ee

and

\be
\nab_a E_{\b}{}^{\uc} - \nab_{\b} E_a{}^{\uc} + T_{a \b}{}^c E_c{}^{\uc}
+ T_{a \b}{}^{\c} E_{\c}{}^{\uc} = 0\ .
\label{43}
\ee

The first of these equations, when projected onto the transverse space, gives

\be
X_{a,b}{}^{c'} = X_{b,a}{}^{c'}\ .
\label{44}
\ee

This also determines $X_{a,\b}{}^{\c'}$ because

\be
X_{a, \b}{}^{\c'}
= -{1 \over 2}(\c^{b})_{\b \c}(\c^{c'})_{j}{}^{k}X_{a,bc'}.
\label{45}
\ee

Furthermore,

\be
X_{a, \b'}{}^{\c} ={1 \over 2}(\c^{b})^{\b \c}(\c^{c'})_{j}{}^{k}X_{a,bc'}.
\ee

Projection onto the worldvolume on the other hand yields

\be
T_{ab}{}^c = -i\L_{b \b}{}^i (\c^c )^{\b \a} \L_{a \a i}.
\label{46}
\ee

Equation \eq{43}, when projected onto the worldvolume with
$(E^{-1})_{\uc}{}^{\c}$,
gives

\be
T_{a \b}{}^{\c} = \L_a{}^{\d'} X_{\b,\d'}{}^{\c} - h_{\b}{}^{\d'}
X_{a,\d'}{}^{\c}.
\label{47}
\ee

The analysis of the projection of \eq{43} onto the transverse space,
however, is more difficult. The resulting equation can be analysed more
easily if we multiply by $m_{a}{}^{b}$; we then find that

\begin{eqnarray}
\nab_{\b j}\tilde{\L}_{b \c k} &=& -{1 \over 2}(\c_{c'})_{jk}
m_{b}{}^{a}m_{d}{}^{e}(\c^{d})_{\b \c}X_{a,e}{}^{c'}
+\e_{jk}(\c_{c})_{\b \c} m_{b}{}^{a}\nab_{a}h^c \nonumber \\
&& -ih_b \tilde{\L}_{c \a j}(\c^{ca})^{\a}{}_{\b}\L_{a \c k}
+ih^a \tilde{\L}^{c}{}_{\a j}(\c_{bc})^{\a}{}_{\b}\L_{a \c k}
-ih_d \tilde{\L}_{c \a j}(\c^{dc})^{\a}{}_{\b}\L_{b \c k} \nonumber \\
&& -{i \over 2}h_c \tilde{\L}_{b \a j}(\c^{dc})^{\a}{}_{\c}\L_{d \b k}
-{i \over 2} h^c \tilde{\L}_{b \c j} \L_{c \b k}
+ih_c \tilde{\L}_{b \a j}(\c^{dc})^{\a}{}_{\b}\L_{d \c k}
+ih^c \tilde{\L}_{b \b j} \L_{c \c k} \nonumber \\
&& +{i \over 2}\e_{jk}h_c \tilde{\L}_{b \a}{}^i (\c^{dc})^{\a}{}_{\c}
\L_{d \b i} +{i \over 2}\e_{jk}h^c \tilde{\L}_{b \c}{}^i \L_{c \b i}.
\label{48}
\end{eqnarray}

Using the fact that $[\nab_A , \nab_B ]h_c = -T_{AB}{}^{D}\nab_{D} h_c
-R_{ABc}{}^{d}h_d$ it can now be shown that

\be
T_{\a \b}{}^{d}\nab_{d}h_{c} = -T_{\a \b}{}^{\c}\nab_{\c}h_c
- \nab_{\a}\nab_{\b}h_c - \nab_{\b}\nab_{\a}h_c
+X_{\a,c}{}^{a'}X_{\b,a'}{}^{d}h_d
+X_{\b ,c}{}^{a'}X_{\a ,a'}{}^{d}h_{d}.
\label{49}
\ee

{}From \eq{49} and \eq{40} it can be seen that $m_b{}^a \nab_a h_c$ is in
fact a function of $h_a$, $\L_b{}^{\b'}$ and
$\nab_{\b j}\tilde{\L}_{b\c k}$. With this substitution \eq{48} can be
rewritten to give a rather
unwieldy expression for $X_{a,e}{}^{c'}$ in terms of
$\nab_{\b j}\tilde{\L}_{b\c k}$ and terms of order $\L^2$.

%%%%%%%%%%%%% Dim 3/2 Torsion Story %%%%%%%%%%%%%%%%%

{\bf Dimension 3/2}

The dimension $3 \over 2$ component of \eq{25} is given by

\be
\nab_a E_{b}{}^{\uc} - \nab_b E_{a}{}^{\uc}
+ T_{ab}{}^{c}E_{c}{}^{\uc} + T_{ab}{}^{\c}E_{\c}{}^{\uc} = 0\ .
\label{50}
\ee

Its projection onto the worldvolume determines $T_{ab}{}^{\c}$ to be

\be
T_{ab}{}^{\c} = \L_{a}{}^{\b'}X_{b,\b'}{}^{\c} - \L_{b}{}^{\b'}X_{a,\b'}{}^{\c}
\label{51}
\ee

while the projection onto the normal space gives

\be
\nab_{[a}\L_{b]}{}^{\c'} = -\L_{[a}{}^{\b'}
X_{b],\b'}{}^{\d}h_{\d}{}^{\c'} - T_{ab}{}^{c}\L_{c}{}^{\c'}
\label{52}
\ee

so that all dimension $3 \over 2$ components are expressible as
functions of lower dimensional quantities. No further components exist
at higher dimensions. The torsion identity \eq{25} therefore
determines all the fields on the worldvolume of the brane off-shell.

\pagebreak

%%%%%%%%%%%%% F Story %%%%%%%%%%%%%%%%%

{\bf The $\cF$ Bianchi Identity}

In addition to the torsion identities all L-branes should satisfy a
further condition which relates superforms on the target space and
worldvolume. In the case of the L5-brane there is a worldvolume 4-form gauge
potential ${\cA}_4$ with corresponding field strength 5-form ${\cF}_5$.
The explicit form for ${\cF}_5$ is

\be
{\cF}_5= d{\cA}_4 -\unC_5
\ee

where $\unC_5$ is the pull-back onto the worldvolume of the target space
5-form potential. The corresonding Bianchi identity is

\be
d{\cF}_5 = - \underline{G}_6
\label{53}
\ee

where $\underline{G_6}$ is the pull-back of $G_6=dC_5$. Equation \eq{53}
can then be solved for ${\cF}_5$. All the components of this tensor are
zero except at dimension zero where we find

\be
{\cF}_{abcde} = -2(m^{-1})_{e}{}^{g}h^{f}\e_{abcdfg}
\label{54}
\ee

where $(m^{-1})_{a}{}^{b}$ is the inverse of $m_{a}{}^{b}$ which is given
explicitly by

\be
(m^{-1})_{a}{}^{b} = {1
\over{1-h^4}}[(1+h^2)\d_{a}{}^{b}-2h_a h^b]\ ,
\ee

where $h^2\equiv h^ah_a$ and $h^4\equiv (h^2)^2$. Furthermore, it is
easy to see that the positive dimension components of the Bianchi
identity are also satisfied. To see this it is convenient to define a
6-form $I_6$ as $I_6 = d{\cF}_5 - \underline{G_{6}}$. It is then
straightforward to show that $dI = 0$ if we remember that the pullback
commutes with the exterior derivative. All components of $I_6$ itself
with more than two spinorial indices must vanish on dimensional grounds.
To show that the other components are also zero we then have to use the
fact that $dI_6=0$ at each dimension independently.
Doing this recursively proves $I_6 = 0$ and thereby establishes \eq{53}.

We conclude this section by noting the relation between the Hodge dual
of ${\cF}_{abcde}$ and $h^a$ which follows from \eq{54}:

\be
{\cF}^a= {2h^a\over 1-h^2}\ , \quad\quad {\cF}^a =
\ft1{5!}\, \e^{abcde} {\cF}_{bcde}\ , \quad\quad h^2 \equiv h^a h_a \ .
\la{fh}
\ee

%%%%%%%%%%%%%%%%%%%%%%%%%%%%%%%%%%%%%%%%%%%%%%%%%%%%%%%%%%%%%%%%%%%%%%%%%%

\section{The Construction of the Action}

%%%%%%%%%%%%%%%%%%%%%%%%%%%%%%%%%%%%%%%%%%%%%%%%%%%%%%%%%%%%%%%%%%%%%%%%%

Recently it was shown how GS-type actions can be systematically constructed for
most branes starting from the superembedding approach \cite{hos}. The only
brane actions that cannot be constructed are those of the $5$-branes in
D=7 and D=11 which both have self-dual anti-symmetric tensors as
components of their supermultiplets. For all other $p$-branes the
starting point for the construction of the action is a closed
$(p+2)$-form, $W_{p+2}$, the Wess-Zumino form, on the worldvolume. This
$(p+2)$-form is given explicitly as the exterior derivative of the (locally
defined) Wess-Zumino potential $(p+1)$-form,
$Z_{p+1}$, $W_{p+2} = dZ_{p+1}$. Since the de Rham cohomology of a
supermanifold is equal to the de Rham cohomolgy of its body, and since
the dimension of the body of the worldvolume superspace is $p+1$, it
follows that $W_{p+2}$ is exact, so that there is a globally defined
$(p+1)$-form $K_{p+1}$ on the worldvolume satisfying

\be
dK_{p+1} = W_{p+2}
\label{55}
\ee

The Green-Schwarz action of the $p$-brane can
then be defined as

\be
S = \int_{M_{o}}L_{p+1}^{0}
\label{56}
\ee

where $M_o$ denotes the (bosonic) body of $M$,

\be
L_{p+1}=K_{p+1}-Z_{p+1}\ ,
\la{lkz}
\ee

and $L_{p+1}^0$ is defined by

\be
L_{p+1}^{0} =dx^{m_{p+1}}\wedge dx^{m_p}\ldots dx^{m_1} L_{m_1\ldots
m_{p+1}}|\ ,
\label{57}
\ee

where the vertical bar indicates evaluation of a superfield at $\th=0$.

By construction $dL_{p+1} = 0$. The $\k$-symmetry of the GS-action is
ensured because, under a worldvolume diffeomorphism generated by a worldvolume
vector field $v$,

\begin{eqnarray}
\d_{v} L_{p+1} &=& {\cal{L}}_{v}L_{p+1} = i_v dL_{p+1} + di_v L_{p+1} \nonumber
\\
&=& d(i_v L_{p+1})\ .
\label{58}
\end{eqnarray}

As explained in \cite{hos}, reparametrisations and $\k$-symmetry
transformations on $M_o$ are essentially the leading components in a
$\th$ expansion of worldvolume diffeomorphisms so that the action given
above is invariant under these transformations by construction.

In the case of the $L5$-brane the Wess-Zumino $7$-form is given by

\be
W_7 = d Z_6=\underline{G_7} + \underline{G_2}{\cF}_5\ .
\label{59}
\ee

where $Z_6$ can be chosen to be

\be
Z_6
= \underline{C_6} + \underline{G_2}{\cA}_4 + \underline{C_1 C_5}\ .
\label{65}
\ee

The globally defined $6$-form $K_6$ needed for the construction of the
action can be solved from

\be
W_7 = dK_6\ .
\label{60}
\ee

The only non-zero component of $K_6$ is the purely bosonic one which is
found to be

\be
K_{abcdef} = -{{1+h^2}\over{1-h^2}}\,\e_{abcdef}.
\label{61}
\ee

Using \eq{fh} we may rewrite the function appearing in \eq{61} as

\be
{{1+h^2}\over{1-h^2}} = \sqrt{1+{\cF}^2}\ ,
\label{62}
\ee

where ${\cF}^2 \equiv {\cF}^a {\cF}_a$. This is the L-brane analogue of
the Dirac-Born-Infeld term in the D-brane action. Therefore $K\equiv
{1\over 6!}\e^{a_1\ldots a_6} K_{a_1\ldots a_6}|$ is given by

\bea
K &=& \sqrt{-g} \sqrt{1+{\cF}^2}\nn\\
  &=& \sqrt { -{\rm det}\ (g_{mn} + {\cF}_m{\cF}_n) }\ ,
\label{63}
\eea

where $g = \det(g_{mn})$ is the determinant of the metric on the
bosonic worldvolume induced by the embedding

\be
g_{mn}={\cE}_m{}^{\una}{\cE}_n{}^{\unb}\h_{\una\unb}
	  =e_m{}^a e_n{}^b \eta_{ab}\ ,
\ee

where

\be
{\cE}_m{}^{\una}= E_m{}^A E_A{}^{\una}|\ ,
\ee

and ${\cF}_a$ is related to ${\cF}_m$ through the worldvolume vielbein
$e_m{}^a$ as ${\cF}m=e_m{}^a {\cF}_a$. The final form for the L5-brane action
is therefore given by

\be
S = \int_{M_{o}} d^6 x\,{\cL}
\ee

where the Green-Schwarz Lagrangian is

\be
{\cL}=\left(\sqrt {-{\rm det}\ (g_{mn} + {\cF}_m{\cF}_n})
- {1\over{6!}}\e^{m_1\ldots m_6} Z_{m_1\ldots m_6}\right)|\ ,
\label{64}
\ee

with  $Z_{m_1\cdots m_6}$ given in \eq{65}.

\section{The Equations of Motion}

{}From the action given in the last section it is straightforward to
derive the equations of motion for the L5-brane. The dynamical variables
in the action are the worldvolume gauge potential $A_{mnpq}$ and the
coordinate $z^{\underline{M}}$ on the target space manifold. The
variation with respect to the worldvolume gauge field is straightforward
and yields

\be
\sqrt{-g} \nabla_m \left({1 \over \sqrt{1+{\cF}^2}}{\cF}^{mnpqr}\right)
= {1 \over 2} \e^{m_1 m_2 npqr}{\cal{E}}_{m_2}{}^{\unM_2}
{\cal{E}}_{m_1}{}^{\unM_1}
G_{\underline{M_1 M_2}}
\label{66}
\ee

where ${\cal{E}}_{m}{}^{\unM} \equiv \del_m Z^{\unM}$, $g$ is again the
induced GS metric on $M_o$ and the covariant derivative is formed using
the Levi-Civita connection of the metric $g$. Note that the Green-Schwarz
embedding matrix (often denoted by $\P$) ${\cE}_m{}^{\unA}$ is given by

\be {\cE}_m{}^{\unA}=\del_m z^{\unM} E_{\unM}{}^{\unA}={\cE}_m{}^{\unM}
E_{\unM}{}^{\unA}\ .
\ee

The
variation of the action with respect to $z^{\underline{M}}$, however, is
rather more involved. Defining $V^{\unM} \equiv \d z^{\unM}$ and
$V^{\unA}=V^{\unM}
E_{\unM}{}^{\unA}$ we find that the
variation of the metric $g_{mn}$ is given by

\be
\d g_{mn} = 2(\del_m V^{\una} + {\cal{E}}_{m}{}^{\ub}
V^{\ua}T_{\underline{\a \b}}{}^{\una}){\cal{E}}_{n \una}.
\label{68}
\ee

Similarly, the variation of ${\cF}_{m_1\cdots m_5}$ gives

\be
\d{\cF}_{m_1\cdots m_5} =
{\cE}_{m_5}{}^{\unA_5} \ldots {\cE}_{m_1}{}^{\unA_1} V^{\unA}
G_{\unA\unA_1\ldots\unA_5}
+ 5 \del_{m_5} \left( V^{\unA_5} {\cE}_{m_4}{}^{\unA_4}
\ldots {\cE}_{m_1}{}^{\unA_1} C_{\unA_1\ldots \unA_5} \right)\ .
\la{70}
\ee

The complete variation of the Green-Schwarz Lagrangian ${\cL}$ (up to
total derivatives) with respect to the $z^{\unM}$ is then given by

\bea
\d{\cL} &=& \sqrt{-g}\,t^{mn}\left(
\partial_{m}V^{\una} -i {\cal{E}}_{m}{}^{\uc}
V^{\ub}(\C^{\una})_{\underline{\b \c}} \right)
{\cal{E}}_{n}{}^{\unb} \eta_{\underline{ab}} \nonumber \\
&+& {1 \over 5!}\sqrt{-g}{1 \over \sqrt{1+{\cF}^2}}{\cF}^{m_1
\ldots m_5}{\cal{E}}_{m_5}{}^{\underline{M}_5}
\ldots {\cal{E}}_{m_1}{}^{\underline{M}_1} V^{\underline{N}}
G_{\underline{N}\underline{M}_1 \ldots \underline{M}_5} \nonumber \\
&-& {1 \over 6!}\e^{m_1 \ldots m_6} {\cal{E}}_{m_6}{}^{\underline{M}_6}
\ldots {\cal{E}}_{m_1}{}^{\underline{M}_1} V^{\underline{N}}
G_{\underline{N}\underline{M}_1 \ldots \underline{M}_6} \nonumber \\
&-& {1 \over 5!} \e^{m_1 \ldots m_6}{\cal{E}}_{m_1}{}^{\underline{M}_1}
V^{\underline{N}} G_{\underline{N}\underline{M}_1}{\cF}_{m_2 \ldots m_6}\ ,
\label{71}
\eea

where we have used \eq{66} and the tensor $t^{mn}$ is given by

\be
t^{mn} = {1 \over \sqrt{1+{\cF}^2}}(g^{mn} + {\cF}^m {\cF}^n)\ .
\label{72}
\ee

It is straightforward to read off the equations of motion from \eq{71}.
For the case of a flat target space one finds, from the vanishing of the
coefficent of $V^{\una}$,

\bea
\nab_m\left( t^{mn} {\cal{E}}_{n}{}^{\unb}
\eta_{\underline{ab}}\right)&=& -{i \over 3!2!}{1 \over
\sqrt{1+{\cF}^2}}{\cF}^{m_1
\ldots m_5}{\cal{E}}_{m_5}{}^{\ub_5} {\cE}_{m_4}{}^{\ub_4}
{\cE}_{m_3}{}^{\unb_3}\ldots {\cal{E}}_{m_1}{}^{\unb_1}
(\C_{\una \unb_1 \ldots \unb_3})_{\ub_4\ub_5} \nonumber \\
&& + {i\over{4!2!}}{\e^{m_1 \ldots m_6}\over \sqrt{-g}}
{\cal{E}}_{m_6}{}^{\ub_6} {\cal{E}}_{m_5}{}^{\ub_5}
{\cE}_{m_4}{}^{\unb_4}\ldots {\cal{E}}_{m_1}{}^{\unb_1}
(\C_{\una \unb_1 \ldots \unb_4})_{\ub_5\ub_6}\ ,
\label{71a}
\eea

where the covariant derivative is again the Levi-Civita derivative with
respect to the GS metric, and, from the vanishing of the coeficient of
$V^{\ua}$,

\bea
t^{mn} {\cal{E}}_{m}{}^{\ub}
(\C^{\una})_{\underline{\b \a}} {\cal{E}}_{n}{}^{\unb}
\eta_{\underline{ab}}&=& -{1 \over 4!}{1 \over \sqrt{1+{\cF}^2}}{\cF}^{m_1
\ldots m_5}{\cal{E}}_{m_5}{}^{\unb_5}
\ldots {\cE}_{m_2}{}^{\unb_2} {\cal{E}}_{m_1}{}^{\ub}
(\C_{\unb_2\ldots\unb_5})_{\underline{\a \b}} \nonumber \\
&& + {1 \over 5!}{\e^{m_1 \ldots m_6}\over \sqrt{-g}}
{\cal{E}}_{m_6}{}^{\unb_6}
\ldots{\cE}_{m_2}{}^{\unb_2} {\cal{E}}_{m_1}{}^{\ub}
(\C_{\unb_2\ldots\unb_6})_{\underline{\a\b}} \nonumber \\
&& +{1 \over 5!} {\e^{m_1 \ldots m_6}\over \sqrt{-g}} {\cal{E}}_{m_1}{}^{\ub}
C_{\underline{\a \b}}{\cF}_{m_2 \ldots m_6}\ ,
\label{71c}
\eea

We shall now compare these equations of motion with the equations of motion one
derives for the L-brane in superspace, i.e. with both worldvolume and target
superspaces. The simplest case to consider is the fermion equation of motion in
a flat target space, equation \eq{71c}, and we shall retrict the discussion to
this example.

The most general Dirac-type equation we can write down in superspace is

\be
M^{ab}E_{b}{}^{\ub}u_a{}^{\una}(\C_{\una} )_{\underline{\b \a}} = 0
\label{73}
\ee

where $M^{ab} = A \eta^{ab} + B h^a h^b$, A and B being scalar
functions of $h_a$. It turns out that this equation (evaluated at $\th=0$) is
equivalent to \eq{71c} provided that we choose the tensor $M_{ab}$ to be equal
to $m_{ab}$. In fact, the equation then reduces to the Dirac equation for the
spinor $\tilde\L$, i.e. the superspace analogue of the linearized
Dirac equation introduced in \eq{41}.

To show that this is the case we need first to evaluate \eq{73} at $\th=0$.
We note that

\be
E_a{}^{\ua}|={\cE}_a{}^{\ub} Q_{\ub}{}^{\ua}
\la{73a}
\ee

where ${\cE}_a{}^{\ua}=e_a{}^m {\cE}_m{}^{\ua}$, $e_a{}^m$ being the inverse
vielbein for the GS metric, and where $Q$ is a projection operator given by

\be
Q_{\ua}{}^{\ub}=(E^{-1})_{\ua}{}^{\c'} E_{\c'}{}^{\ub}|\ .
\la{73b}
\ee

It is easy to evaluate $Q$ explicitly; one finds

\be
2Q=1 + {1\over 6!}\e^{a_1\ldots a_6} \hat\C_{a_1\ldots a_6} -h^a\hat\C_a
-{1\over5!} \e^{a b_1\ldots b_5} h_a \hat\C_{b_1\ldots b_5}\ ,
\la{73c}
\ee

where

\be
\hat\C_a :={\cE}_a{}^{\una}\C_{\una}
\la{73d}\ .
\ee

To show the equivalence of the two  fermionic equations one simply computes
\eq{73} and then right multiplies it by $2(1-h^4)^{-1}(1-h^a\hat\C_a)$. The
resulting equation then has the same form as equation \eq{71c} when this is
expressed in terms of $h$ rather than ${\cF}$. This equation takes the form

\bea
\left({1-h^2\over 1+h^2} \h^{ab} + 4{h^a h^b\over
1-h^4}\right){\cE}_b{}^{\unb}(\hat\C_a)_{\ua\ub} &=& -{1\over4!}{2\over
1-h^4}\e^{ba_1\ldots a_5} h_{a_5}{\cE}_b{}^{\ub}(\hat\C)_{a_1\ldots
a_4})_{\ua\ub}\nonumber \\
&&-{1\over5!} \e^{b a_1\ldots a_5} {\cE}_b{}^{\ub}(\hat\C_{a_1\ldots
a_5})_{\a\b}- {2h^b\over 1-h^2}{\cE}_b{}^{\ub} C_{\ua\ub}\ .
\la{73e}
\eea

One might wonder whether other choices of the tensor $M_{ab}$ could lead to a
different consistent set of equations of motion. Although we have not checked
this we believe that it is unlikely. In other words, if one were to make a
different choice for $M_{ab}$ one would find non-linear inconsistencies at
higher dimension arising as a consequence.

%%%%%%%%%%%%%%%%%%%%%%%%%%%%%%%%%%%%%%%%%%%%%%%%%%%%%%%%%%%%%%%%%%%%

\section{L-branes in D=7 and D=8}

%%%%%%%%%%%%%%%%%%%%%%%%%%%%%%%%%%%%%%%%%%%%%%%%%%%%%%%%%%%%%%%%%%%%

The L3-brane in $D=7$ and the L$4$-brane in $D=8$ can in principle be
derived by double dimensional reduction from the L5-brane in $D=9$.
However, it is simpler to construct them directly using the same
techniques that were used for the L5-brane. To derive the action we only
need to analyse the torsion and Bianchi identities at dimension zero as
this information is sufficient to compute the $(p+1)$-form $K_{p+1}$
which, together with the Wess-Zumino form $Z_{p+1}$ determines the
action from \eq{lkz} and \eq{56}.

%%%%%%%%%%%%%%%%%%%%%%%%%%%%%%%%%%%%%%%%%%%%%

\subsection{The L$4$-Brane in $D=8$}

%%%%%%%%%%%%%%%%%%%%%%%%%%%%%%%%%%%%%%%%%%%%%

The analysis of the dimension zero torsion identity is similar to the
L5-brane case; from \eq{12'} one again finds the constraint \eq{15} on
the field $h_{\a}{}^{\b'}\rightarrow h_{\a i}{}^{\b j}$ which is solved
by

\be
h_{\a i\b j}=\e_{ij}\left( C_{\a\b} S + (\c^a)_{\a\b}\, h_a \right)\ .
\la{p4}
\ee

The scalar field $S$ can be identified with the auxiliary field of the
linear multiplet in $d=5$. The dimension 0 components of the
worldvolume torsion are found to be

\be
T_{\a i,\b j}{}^a = -i\e_{ij}\left((\c^b)_{\a\b} m_b{}^a \pl C_{\a\b}
m^a\right)\ ,
\la{79}
\ee

where

\bea
m_b{}^a &=& (1 - h^2 + S^2)\d_b{}^a \pl 2h^a h_b\ ,\\
m_a &=& 2S\,h_a\ . \la{m's}
\eea

{}From the Bianchi identity

\be
d{\cF}_4=-{\unG}_5\ ,
\ee

it follows that

\be
{\cF}_a = -{2h_a\over 5(1-h^2+S^2)}\ , \la{fh4}
\ee

where ${\cF}_{abcd}= \e_{abcde} {\cF}^e$.

To construct an action, we need to consider the Wess-Zumino form

\be
W_6 = d Z_5=\underline{G_6} + \underline{G_2}{\cF}_4\ .
\label{59a}
\ee

with  $Z_5$ given by

\be
Z_5
= \underline{C_5} + \underline{G_2}{\cA}_3 + \underline{C_1 C_4}\ .
\label{65a}
\ee

The globaly defined $5$-form $K_5$ needed for the construction of the
action can be solved from $W_6= dK_5$. We find that the only non-zero

component of $K_5$ is the purely bosonic one given by

\be
K_{abcde} = -\left( {1+h^2-S^2\over 1-h^2+S^2}\right)\,\e_{abcde}.
\label{61a}
\ee

Using the action formula $ S = \int_{M_{o}} d^5x\,{\cL} $ where
${\cL}=K_5-Z_5 $ and recalling \eq{fh4} and \eq{61a} we find that the
Lagrangian can be written as

\be
{\cL}= \left( \sqrt{-{\rm det}\, \left(g_{mn} + 25(1-S^2){\cF}_m{\cF}_n
\right)}
- {1\over{5!}}\e^{m_1\ldots m_5} Z_{m_1\ldots m_5}\right)|\ ,
\ee

with $Z_{m_1\cdots m_5}$ given in \eq{65a}.

%%%%%%%%%%%%%%%%%%%%%%%%%%%%%%%%%%%%%%%%%%%%%%%%

\subsection{The L$3$-Brane in $D=7$ }

%%%%%%%%%%%%%%%%%%%%%%%%%%%%%%%%%%%%%%%%%%%%%%%%

The construction of the L$3$-brane action in $D=7$ parallels exactly the
constructions presented above. We find that the analogues of the equations
\eq{p4}-\eq{65a} for this case are

\be
h_{\a i\b j}=\e_{ij}\left(C_{\a\b} S + (\c_5)_{\a\b} T +
(\c_5\c^a)_{\a\b}\, h_a\right)\ ,
\la{p3}
\ee

where $S,T$ are the auxilary fields, and

\be
T_{\a i, \b j}{}^a = -i\e_{ij}\left((\c^b)_{\a\b} m_b{}^a \pl
                      (\c^{ba})_{\a\b} m_b\right)\ ,
\la{80}
\ee

where

\bea
m_a{}^b &=& (1 -h^2+S^2+T^2)\d_b{}^a \pl 2h^a h_b,\\
m_a &=& 2T h_a\ .
\eea

Furthermore, starting from the Bianchi identity

\be
d{\cF}_3=-{\unG}_4\ ,
\ee

we find that

\be
{\cF}_a = -{h_a\over 3(1-h^2-S^2-T^2)}\ , \la{fh3}
\ee

where ${\cF}_{abc}= \e_{abcd} {\cF}^d$ and that

\be
W_5 = d Z_4=\underline{G_5} + \underline{G_2}{\cF}_3\ .
\label{59b}
\ee

\be
Z_4
= \underline{C_4} + \underline{G_2}{\cA}_2 + \underline{C_1 C_3}\ .
\label{65b}
\ee

\be
K_{abcd} = -\left( {1+h^2-S^2-T^2\over 1-h^2+S^2+T^2}\right) \,\e_{abcd}.
\label{61b}
\ee

Again, using the action formula $ S = \int_{M_{o}} d^4 x\,{\cL} $ where
${\cL}=K_4-Z_4 $ and recalling \eq{fh3} and \eq{61b} we find that the
Lagrangian for the L$3$-brane in $D=7$ can be written as

\be
{\cL}= \left(\sqrt{-{\rm det}\,\left(g_{mn} + 36(1-S^2-T^2){\cF}_m{\cF}_n
\right)}
- {1\over{4!}}\e^{m_1\ldots m_4} Z_{m_1\ldots m_4}\right)|\ ,
\ee

with $Z_{m_1\cdots m_4}$ given in \eq{65b}.

%%%%%%%%%%%%%%%%%%%%%%%%%%%%%%%%%%%%%%%%%%%%%%%%%%%%%%%%%%%%%%%

\section{Comments}

%%%%%%%%%%%%%%%%%%%%%%%%%%%%%%%%%%%%%%%%%%%%%%%%%%%%%%%%%%%%%%%

We have seen that L-branes are examples of a class of $p$-branes with an
unconventional worldvolume supermultiplet, namely the linear
multiplet. These branes arise naturally within the superembedding
approach but have so far been neglected in the literature. In marked difference
to most other branes we have seen
that the linear multiplets are off-shell. One of the consequences is that the
usual torsion
equations are not the equations of motion for the branes. In fact these
have to be derived from an action. We have illustrated the dynamics of
the L$5$-brane by solving the highly non-linear torsion equations in
terms of a divergence-free vector $h_a$ and a vector-spinor
$\L_{a}{}^{\a'}$ in a flat target space background. For the L$4$-brane in
$D=8$, an aditional auxiliary scalar $S$ and for the L$3$-brane in $D=7$, the
additional auxiliary scalars $(S,T)$ were shown to arise. Using a general
action principle which is valid for the
construction of actions for most branes we have found the Green-Schwarz
action of the L-branes. For the L$5$-brane we derived the Green-Schwarz
equations of motion and we illustrated the relationship between the
equations of motion in superspace and those derived from the action in
the case of the spinor equation.

We have noted that the L$5$-brane can be viewed as the dimensional reduction of
a superfivebrane in $D=10$ dimensionally reduced to $D=9$, followed by
dualization of the scalar corresponding to the extra dimension to a 4-form
potential. This relation between a fivebrane in $D=10$ and L$5$-brane in $D=9$
is similar to the relation between the M$2$-brane in $D=11$ and D$2$-brane in
$D=10$. The latter relation has been called M-duality \cite{pkt4} which relates
Type IIA string theory to M-theory in the strong coupling limit. Other
worldvolume duality transformations have also been studied. Indeed, the
worldvolume $U(1)$ gauge fields arising in D$p$-branes have been dualized to
$(p-2)$-form gauge fields for $p \le 4$. We refer the reader to \cite{jhs4} for
various aspects of these dualizations and for an extensive list of references
for earlier works on the subject. The point we wish to emphasize here is that,
while the methods employed in the literature so far for worldvolume
dualizations become forbiddingly complicated beyond $p=4$, the superembedding
approach provides an alternative and simpler method which seems to apply
universally to all possible branes. Regardless of the approach taken, results
obtained in the area of worldvolume dualizations are hoped to provide further
connections among a large class of branes that arise in the big picture of
M-theory.

As mentioned earlier, we have focused our attention on flat target
superspaces in this paper. The generalisation of our work to curved target
superspace is straightforward. Consider the case of the L$5$-brane, for
example. The results presented in sections 4 and 5 remain the same in
curved superspace. The superforms occurring in the Wess-Zumino term
\eq{65}, however, now live in a curved target superspace. The field
strength forms $G_2,G_6,G_7$ still obey the Bianchi identities \eq{ggg},
but they will have more nonvanishing components than those given in
\eq{10}. The expected solution to the full set of Bianchi identities is
$N=1,D=9$ supergravity coupled to a single vector multiplet, as
resulting from the dimensional reduction of the $N=1,D=10$ supergravity
theory in its dual formulation. Thus, the $D=9$ field content is

\bea
&& {\rm Supergravity:}  \quad\quad
(g_{mn}, C_{m_1\cdots m_5},\,C_{m_1\cdots m_6},\, \phi)\,
(\psi_m,\, \chi)  \nn\\
&& {\rm Maxwell}: \quad\quad\quad \ \ (C_m, \s)\, (\l)
\eea

where we have grouped the bosonic and fermionic fields separately, in a self
explanatory notation. It should be noted that the fields $C_5$ and $C_6$ come
from the dimensional reduction of a 6-form potential, and $(C_1,\s)$ come from
the Kaluza-Klein reduction of the metric in $D=10$.  The coupling of $n$ vector
multiplets to $N=1,D=9$ supergravity has been determined in \cite{gns} in a
formalism that contains $(B_2,B_1)$ which are the duals of $(C_5,C_6)$ and in
\cite{at} in a formalism which has the fields $(C_5,B_1)$.  It would be
interesting to study the brane solitons of these theories and to determine the
maximum possible symmetries they may exhibit. It is known that the 5-brane
solution of $N=1,D=10$ supergravity involves a nonconstant dilaton and
consequently it does not give rise to an $AdS_7 \times S^3$ geometry in the
near horizon limit (see \cite{duff1} for a detailed discussion of this matter
and for earlier references). On the other hand, there exists a singleton field
theory which is a candidate for the description of a fivebrane in this
background \cite{nst}. Given the close relation between the L$5$-brane in $D=9$
and the 5-brane in $D=10$, it is natural to study the L$5$-brane solution of
the Einstein-Maxwell supergravity in $D=9$ and to determine if it permits a
constant dilaton, thereby possibly giving rise to an $AdS_7 \times S^2$ near
horizon geometry. It should be noted that the candidate singleton field theory
in this case would be characterized by a a superconformal field theory of a
free linear supermultiplet in six dimensions discussed in section 3, and whose
superconformal transformations can be found in \cite{bps}.

The field content of the target space Einstein-Maxwell supergravities
relevant to the L$4$-brane in $D=8$ and L$3$-brane in $D=7$ remains to be
worked out in detail as well. The expected results are various versions
of Einstein-Maxwell supergravities in which certain fields have been
dualized.

\bigskip

\section*{Acknowledments}

We thank M.J. Duff for useful discussions. This research has been
supported in part by NSF Grant PHY-9722090.

\bigskip

\section*{Appendix}

\subsection{Superspace Conventions}

\subsection*{D=9 $\to$ d=6}

The superembeddings studied in this note are taken to break half of the
original target space supersymmetries. This implies that the
worldvolume of a brane has $1 \over 2$ the number of fermionic coordinates of
the target space. In the example of the L5-brane this means that the
target space $\unM$ has $9$ bosonic and $16$ fermionic coordinates while
the worldvolume of the brane $M$ has $6$ bosonic and $8$ fermionic
coordinates. To study the embedding in more detail it is convenient to
split the $9$-dimensional $\c$-matrices in a way that reflects the
embedding, i.e. $SO(1,8) \rightarrow SO(1,5) \xz SU(2)$, as follows:

\begin{eqnarray}
(\C^{a})_{\ua}{}^{\ub} & = & \pmatrix{
\bf 0 &(\c^{a})_{\a \b} \cr
(\c^{a})^{\a \b} &\bf 0}\d_{i}{}^j \nonumber \\
(\C^{a'})_{\ua}{}^{\ub} & = & \pmatrix{
\d_{\a}{}^{\b} &\bf 0 \cr
\bf 0 &-\d^{\a}{}_{\b}}(\c^{a'})_{ij}.
\label{97}
\end{eqnarray}

The $9$-dimensional charge conjugation matrix $C_{\underline{\a \b}}$ is given
by

\be
C_{\underline{\a \b}} = \e_{ij} \pmatrix{
\bf 0 &-\d_\a{}^\b \cr
\d^\a{}_\b &\bf 0}.
\label{98}
\ee

The conventions for the spinors are chosen such that $9$-dimensional
spinors $\psi_{\ua}$ are pseudo-Majorana while $6$-dimensional spinors
are chosen to be symplectic Majorana-Weyl. This implies that

\be
\psi_{\ua} \rightarrow \left\{ \matrix{
\psi_\a  & = & \psi_{\a i} \cr
\psi_{\a'} & = & \psi_i^{\a}.} \right.
\label{99}
\ee

$SU(2)$-indices can be raised and lowered with $\e_{ij}$ using the
convention that $\l_i = \l^j \e_{ji}$. A useful equation for
manipulations of $SU(2)$ $\c$-matrices is

\be
(\c^{c'})_{jk}(\c_{c'})_{il} = \e_{ij}\e_{kl}+\e_{jl}\e_{ik}.
\label{100}
\ee

The conversion between spinor and Lorentz indices is governed by

\be
K_{\b}{}^{\c} = {1 \over 4} \d_j{}^k (\c^{bc})_{\b}{}^{\c} K_{bc}
+ \d_{\b}{}^{\c} K_{jk}
\label{102}
\ee

and

\be
K_{\b}{}^{\c'} = -{1 \over 2}(\c^{b})_{\b \c}(\c^{c'})_j{}^k K_{bc'}.
\label{103}
\ee

\subsection*{D=8 $\to$ d=5}

\be
C_ {\underline{\a \b}}=\e_{ij} \pmatrix{
C_{\a \b}  & \bf 0  \cr
\bf 0  & -C_{\a\b} },
\la{104}
\ee

\begin{eqnarray}
(\C^{a'})_{\ua}{}^{\ub} & = & (\c^{a'})_i{}^j \pmatrix{
\bf  0 & \d_{\a}{}^{\b} \cr
{\d}^{\a}{}_{\b} & \bf  0 }  \nonumber \\
(\C^{a})_{\ua}{}^{\ub} & = & \d_i{}^j \pmatrix {
(\c^a)_{\a}{}^{\b} & \bf 0 \cr
\bf 0 & - (\c^a)^{\a}{}_{\b} } ,
\la{105}
\end{eqnarray}

\be
\psi_{\ua} \rightarrow \left\{ \matrix{
\psi_\a  & = & \psi_{\a i} \cr
\psi_{\a'} & = &\phi_{\a i} .} \right.
\la{106}
\ee

\subsection*{D=7 $\to$ d=4}

\bea
C_ {\underline{\a \b}}=\e_{ij} \pmatrix{
\bf 0  & - C_{\a \b} \cr
C^{\a\b} &\bf 0  },
\la{107}
\eea

\begin{eqnarray}
(\C^{a'})_{\ua}{}^{\ub} & = & -(\c^{a'})_{ij} \pmatrix{
\d_{\a}{}^{\b} & \bf  0 \cr
\bf  0 & -{\d}_{a}{}^{\b} }  \nonumber \\
(\C^{a})_{\ua}{}^{\ub} & = & \d_i{}^j \pmatrix {
\bf 0 &(\c^a)_{\a}{}^{\b} \cr
(\c^a)_{\a}{}^{\b} & \bf 0 } ,
\la{108}
\end{eqnarray}

\be
\psi_{\ua} \rightarrow \left\{ \matrix{
\psi_\a  & = & \psi^{\pl}_{\a i} \cr
\psi_{\a'} & = &\psi^{-}_{\a i} .} \right.
\la{109}
\ee

Our conventions for superforms are as follows. Gauge potentials $A_q$ are
defined by

\be
A_q = {1 \over q!} dz^{M_q} \ldots dz^{M_1} A_{M_1 \ldots M_q}
\label{110}
\ee

and field strengths by

\be
F_q = {1 \over (q+1)!} dz^{M_{(q+1)}} \ldots dz^{m_1} F_{M_1 \ldots M_{(q+1)}}.
\label{111}
\ee

Wedge products between forms are understood. The exterior derivative $d$
acting on a $q$-form $\rho$ we define by

\be
d \rho = dz^{M_{(q+1)}} \ldots dz^{M_1} \partial_{M_1} \rho_{M_2 \ldots
M_{(q+1)}}
\label{112}
\ee

and the interior product $i_v$ by

\be
i_v \rho = q dz^{M_q} \ldots dz^{M_2} v^{M_1} \rho_{M_1 \ldots M_{(q+1)}}.
\label{113}
\ee

Both act from the right on products of forms, i.e.

\be
d( \omega \rho) = \omega (d \rho) + (-1)^q (d \omega) \rho
\label{114}
\ee

and

\be
i_v (\omega \rho) = \omega (i_v \rho) + (-1)^q (i_v \omega ) \rho
\label{115}
\ee

where $\omega$ is a $p$-form and $\rho$ is a $q$-form.

\end{document}